\documentclass[aip,jcp,reprint,twocolumn,showpacs,superscriptaddress]{revtex4}

\pdfoutput=1

\usepackage{amsmath,amssymb}
\usepackage{bm}
\usepackage{graphicx}

\newcommand{\vm}{\bm{m}}
\newcommand{\xv}{\widehat{{x}}}
\newcommand{\yv}{\widehat{{y}}}
\newcommand{\zv}{\widehat{{z}}}
\newcommand{\vR}{\bm{R}}
\newcommand{\parallelsum}{\, \mathbin{\!/\mkern-5mu/\!} \,}

\newcommand{\NAT}{\textrm{NAT}}
\newcommand{\AT}{\textrm{AT}}

\usepackage{ulem}
\usepackage{color}

\newcommand{\AM}[1]{\textcolor{black}{#1}}

\begin{document}

{\it\small
\noindent Journal Reference: J.~Chem.~Phys.~\textbf{141}, 124904 (2014)\\
URL: http://scitation.aip.org/content/aip/journal/jcp/141/12/10.1063/1.4896147\\
DOI: 10.1063/1.4896147\\[1.cm]
}

$\mbox{\hspace{1.cm}}$
\vspace{.9cm}

\title{
Structural control of elastic moduli in ferrogels and the importance of non-affine deformations
}

\author{Giorgio Pessot}
\affiliation{Institut f\"ur Theoretische Physik II: Weiche Materie, Heinrich-Heine-Universit\"at D\"usseldorf, D-40225 D\"usseldorf, Germany}

\author{Peet Cremer}
\affiliation{Institut f\"ur Theoretische Physik II: Weiche Materie, Heinrich-Heine-Universit\"at D\"usseldorf, D-40225 D\"usseldorf, Germany}

\author{Dmitry Y.\ Borin}
\affiliation{Technische Universit\"at Dresden, Institute of Fluid Mechanics, D-01062, Dresden, Germany}

\author{Stefan Odenbach}
\affiliation{Technische Universit\"at Dresden, Institute of Fluid Mechanics, D-01062, Dresden, Germany}

\author{Hartmut L\"owen}
\affiliation{Institut f\"ur Theoretische Physik II: Weiche Materie, Heinrich-Heine-Universit\"at D\"usseldorf, D-40225 D\"usseldorf, Germany}

\author{Andreas M.\ Menzel}
\email{menzel@thphy.uni-duesseldorf.de}
\affiliation{Institut f\"ur Theoretische Physik II: Weiche Materie, Heinrich-Heine-Universit\"at D\"usseldorf, D-40225 D\"usseldorf, Germany}

\date{\today}

\begin{abstract}
One of the central appealing properties of magnetic gels and elastomers is that their elastic moduli
 can reversibly be adjusted from outside by applying magnetic fields. The impact of the internal magnetic
 particle distribution on this effect has been outlined and analyzed theoretically.
In most cases, however, affine sample deformations are studied and often regular particle arrangements
 are considered.
Here we challenge these two major simplifications by a systematic approach using a minimal dipole-spring model. 
Starting from different regular lattices, we take into account increasingly randomized structures,
 until we finally investigate an irregular texture taken from a real experimental sample.
On the one hand, we find that the elastic tunability qualitatively depends on the structural properties,
 here in two spatial dimensions.
On the other hand, we demonstrate that the assumption of affine deformations leads to increasingly erroneous
 results the more realistic the particle distribution becomes. 
Understanding the consequences of the assumptions made in the modeling process is important on our way to support an improved design of these fascinating materials. 
\end{abstract}



\maketitle


\section{Introduction} 

In the search of new materials of outstanding novel properties, one route is to combine the features
 of different compounds into one composite substance \cite{jarkova2003hydrodynamics,ajayan2000single,
 stankovich2006graphene,glotzer2007anisotropy,yuan2011one}. 
Ferrogels and magnetic elastomers provide an excellent example for this approach. 
They consist of superparamagnetic or ferromagnetic particles of nano- or micrometer size embedded
 in a crosslinked polymer matrix \cite{filipcsei2007magnetic}.
In this way, they combine the properties of ferrofluids and magnetorheological fluids
 \cite{klapp2005dipolar,rosensweig1985ferrohydrodynamics,odenbach2003ferrofluids,odenbach2003magnetoviscous,
huke2004magnetic,odenbach2004recent,ilg2005structure, embs2006measuring,ilg2006structure,gollwitzer2007surface}
 with those of conventional polymers and rubbers \cite{strobl1997physics}:
 we obtain elastic solids, the shape and mechanical properties of which can be \AM{changed} reversibly
 from outside by applying external magnetic fields \cite{zrinyi1996deformation,deng2006development,
stepanov2007effect, filipcsei2007magnetic,guan2008magnetostrictive,bose2009magnetorheological,
gong2012full,evans2012highly,borin2013tuning}. 

This magneto-mechanical coupling opens the door to a multitude of applications. 
Deformations induced by external magnetic fields suggest a use of the materials as soft actuators
 \cite{zimmermann2006modelling} or as sensors to detect magnetic fields and field gradients
 \cite{szabo1998shape,ramanujan2006mechanical}. 
The non-invasive tunability of the mechanical properties by external magnetic fields makes them candidates
 for the development of novel damping devices \cite{sun2008study} and vibration absorbers
 \cite{deng2006development} that adjust to changed environmental conditions. 
Finally, local heating due to hysteretic remagnetization losses in an alternating external magnetic field
 can be achieved. This effect can be exploited in hyperthermal cancer treatment
 \cite{babincova2001superparamagnetic,lao2004magnetic}. 

In recent years, several theoretical studies were performed to elucidate the role of the spatial magnetic
 particle distribution on these phenomena \cite{ivaneyko2011magneto,wood2011modeling,camp2011effects,
stolbov2011modelling, ivaneyko2012effects,weeber2012deformation,gong2012full, zubarev2012theory,
zubarev2013effect, zubarev2013magnetodeformation,han2013field,ivaneyko2014mechanical}.
It turns out that the particle arrangement has an even qualitative impact on the effect that external
 magnetic fields have on ferrogels.
That is, the particle distribution within the samples determines whether the systems elongate
 or shrink along an external magnetic field, or whether an elastic modulus increases or decreases
 when a magnetic field is applied.
As a first step, many of the theoretical investigations focused on regular lattice structures
 of the magnetic particle arrangement \cite{ivaneyko2011magneto,ivaneyko2012effects,ivaneyko2014mechanical}.
Meanwhile, it has been pointed out that a touching or clustering of the magnetic particles
 and spatial inhomogeneities in the particle distributions can have a major influence
 \cite{stolbov2011modelling,gong2012full,annunziata2013hardening,zubarev2013effect,
zubarev2013magnetodeformation,han2013field}.
More randomized or ``frozen-in'' gas-like distributions were investigated \cite{wood2011modeling,
camp2011effects,stolbov2011modelling,gong2012full, zubarev2012theory,zubarev2013magnetodeformation}.
Yet, typically in these studies an affine deformation of the whole sample is assumed,
 i.e.\ the overall macroscopic deformation of the sample is mapped uniformly to all distances in the system.
An exception is given by microscopic \cite{weeber2012deformation} and finite-element studies
 \cite{stolbov2011modelling,gong2012full,han2013field}, but the possible implication of the assumption
 of an affine deformation for non-aggregated particles remains unclear from these investigations. 

Here, we systematically challenge these issues using the example of the compressive elastic modulus under
 varying external magnetic fields. We start from regular lattice structures that are more and more randomized.
In each case, the results for affine and non-affine deformations are compared.
Finally we consider a particle distribution that has been extracted from the investigation of a real
 experimental sample.
It turns out that the assumption of affine deformations growingly leads to erroneous results with
 increasingly randomized particle arrangements and is highly problematic for realistic particle distributions. 

In the following, we first introduce our minimal dipole-spring model used for our investigations.
We then consider different lattice structures: rectangular, hexagonal, and honey-comb, all of them with
 increasing randomization. Different directions of magnetization are taken into account.
Finally, an irregular particle distribution extracted from a real experimental sample is considered,
 before we summarize our conclusions. 


\section{Dipole-Spring Minimal Model}

For reasons of illustration and computational economics, we will work with point-like particles
 confined in a two-dimensional plane with open boundary conditions.
\AM{On the one hand, we will study regular lattices, for which simple analytical arguments can be given
 to predict whether the elastic modulus will increase or decrease with increasing magnetic interaction.
These lattices will also be investigated after randomly introducing positional irregularities.
Such structures could reflect the properties of more realistic systems, for example those of thin regularly
 patterned magnetic block-copolymer films \cite{garcia2003mesoporous,kao2013toward}.}

On the other hand, irregular particle distributions in a plane to some extent reflect the situation in
 three dimensional anisotropic
 magnetic gels and elastomers \cite{collin2003frozen,varga2003smart,bohlius2004macroscopic,
 gunther2012xray,borbath2012xmuct,gundermann2013comparison}.
In fact, our example of irregular particle distribution is extracted from a real anisotropic experimental sample.
These anisotropic materials are manufactured under the presence of a strong homogeneous external magnetic field.
It can lead to the formation of chain-like particle aggregates that are then ``locked-in'' during the final
 crosslinking procedure.
These chains lie parallel to each other along the field direction and can span the whole sample
 \cite{gunther2012xray}.
To some extent, the properties in the plane perpendicular to the anisotropy direction may be represented by
 considering the two-dimensional cross-sectional layers on which, in this work, we will focus our attention.

Our system is made of $N=N_x\times N_y$ point-like particles with positions $\vR_i$, $i=1\dots N$,
 each carrying an identical magnetic moment $\vm$.
That is, we consider an equal magnetic moment induced for instance by an external magnetic field
 in the case of paramagnetic particles, or an equal magnetic moment of ferromagnetic particles
 aligned along one common direction.
\AM{We assume materials in which the magnetic particles are confined in pockets of the polymer mesh.
They cannot be displaced with respect to the enclosing polymer matrix, i.e.~out of their
 pocket locations.}
Neighboring particles are coupled by springs of different unstrained length $l_{ij}^0$ according
 to the selected initial particle distribution.
All springs have the same elastic constant $k$.
\AM{The polymer matrix, represented by the springs, is assumed to have a vanishing magnetic susceptibility.
Therefore it does not directly interact with magnetic fields.
(The reaction of composite bilayered elastomers of non-vanishing
 magnetic susceptibility to external magnetic fields was investigated recently in a different study \cite{allahyarov2014magnetomechanical}).}

The total energy $U$ of the system is the sum of elastic and magnetic energies
 \cite{annunziata2013hardening,cerda2013phase,sanchez2013filaments} $U_{el}$ and $U_m$ defined by
\begin{equation}\label{eel}
 U_{el}= \frac{k}{2}\sum_{\langle ij\rangle} {\left( r_{ij} -l_{ij}^0 \right)}^2,
\end{equation}
where $\langle ij\rangle$ means sum over all the couples connected by springs, $\bm{r}_{ij}=\vR_j -\vR_i$,
 $r_{ij}=|\bm{r}_{ij}|$ and
\begin{equation}\label{emagn}
 U_{m}= \frac{\mu_0 m^2}{4\pi}\sum_{i<j} \frac{r_{ij}^2 -3{(\widehat{\vm}\cdot\bm{r}_{ij})}^2}{r_{ij}^5},
\end{equation}
where $i<j$ means sum over all different couples of particles, and $\widehat{\vm}=\vm/m$
 is the unit vector along the direction of $\vm$.
In our reduced units we measure lengths in multiples of $l_0$ and energies in multiples of
$k {l_0}^2$; here we define $l_0=1/\sqrt{\rho}$, where $\rho$ is the particle area density.
\AM{To allow a comparison between the different lattices we choose the initial density always the
 same in each case.}
Furthermore, our magnetic moment is measured in multiples of $m_0=\sqrt{4\pi k^2 {l_0}^5/\mu_0}$.

Estimative calculations show that the magnetic moments obtainable in real systems are $4-5$ orders
 of magnitude smaller than our reduced unit for the magnetic moment, so only the behavior for the
 rescaled $|\vm|/m_0=m/m_0\ll 1$ would need to be considered.
Here, we run our calculations for $m$ as big as possible, until the magnetic forces become so strong
 as to cause the lattice to collapse, which typically occurs beyond realistic values of $m$.
\AM{After rescaling, the magnetic moment $\vm$ is the only remaining parameter in our equations
 which can be used to tune the system for a given particle distribution.}


\section{Elastic Modulus from Affine and Non-Affine Transformations}
We are interested in the elastic modulus $E$ for dilative and compressive deformations of the system,
 as a function of varying magnetic moment and lattices of different orientations and particle arrangements.
For a fixed geometry and magnetic moment $\vm$, once we have found the equilibrium state of
 minimum energy of the system, we calculate $E$ as the second derivative of total energy with respect to
 a small expansion/shrinking of the system, here in $x$-direction:
\begin{equation}\label{elmod}
 E=\frac{d^2 U}{{d\delta_x}^2} \simeq \frac{U(-\delta_x)+U(\delta_x)-2U(0)}{{\delta_x}^2}.
\end{equation}
$\delta_x$ is a small imposed variation of the sample length along $\xv$.
\AM{In order to remain in the linear elasticity regime, $\delta_x$ must imply an elongation
 of every single spring by a quantity small compared to its unstrained length.
In our calculations we chose a total length change of the sample of $\delta_x=L_x/100\sqrt{N}\simeq
 l_{0}/100$ throughout, where $L_x$ is the equilibrium length of the sample along $\xv$.
Thus, on average, each spring is strained along $\xv$ by less than $1\%$.
To indicate the direction of the induced strain, we use the letter $\varepsilon$ in the figures below.
The magnitude of the strain follows as $|\varepsilon|=\delta_x/L_x\simeq 10^{-4}-10^{-3}$. Strains of such magnitude were for example applied experimentally using a piezo-rheometer \cite{collin2003frozen}.}
A natural unit to measure the elastic modulus $E$ in Eq.~(\ref{elmod}) is given by the elastic
 spring constant $k$. 

There are different ways of deforming the lattices in order to find the equilibrium configuration
 of the system and calculate the elastic modulus. We will demonstrate that considering non-affine
 instead of affine transformations can lead to serious differences in the results, especially
 for randomized and realistic particle distributions. 

An affine transformation (AT) conserves parallelism between lines and in each direction
 modifies all distances by a certain ratio.
In our case of a given strain in $x$-direction, in AT we obtain the equilibrium state by minimizing
 the energy over the ratio of compression/expansion in $y$-direction.

\begin{figure}
  \begin{center}
    \includegraphics[width=5.5cm]{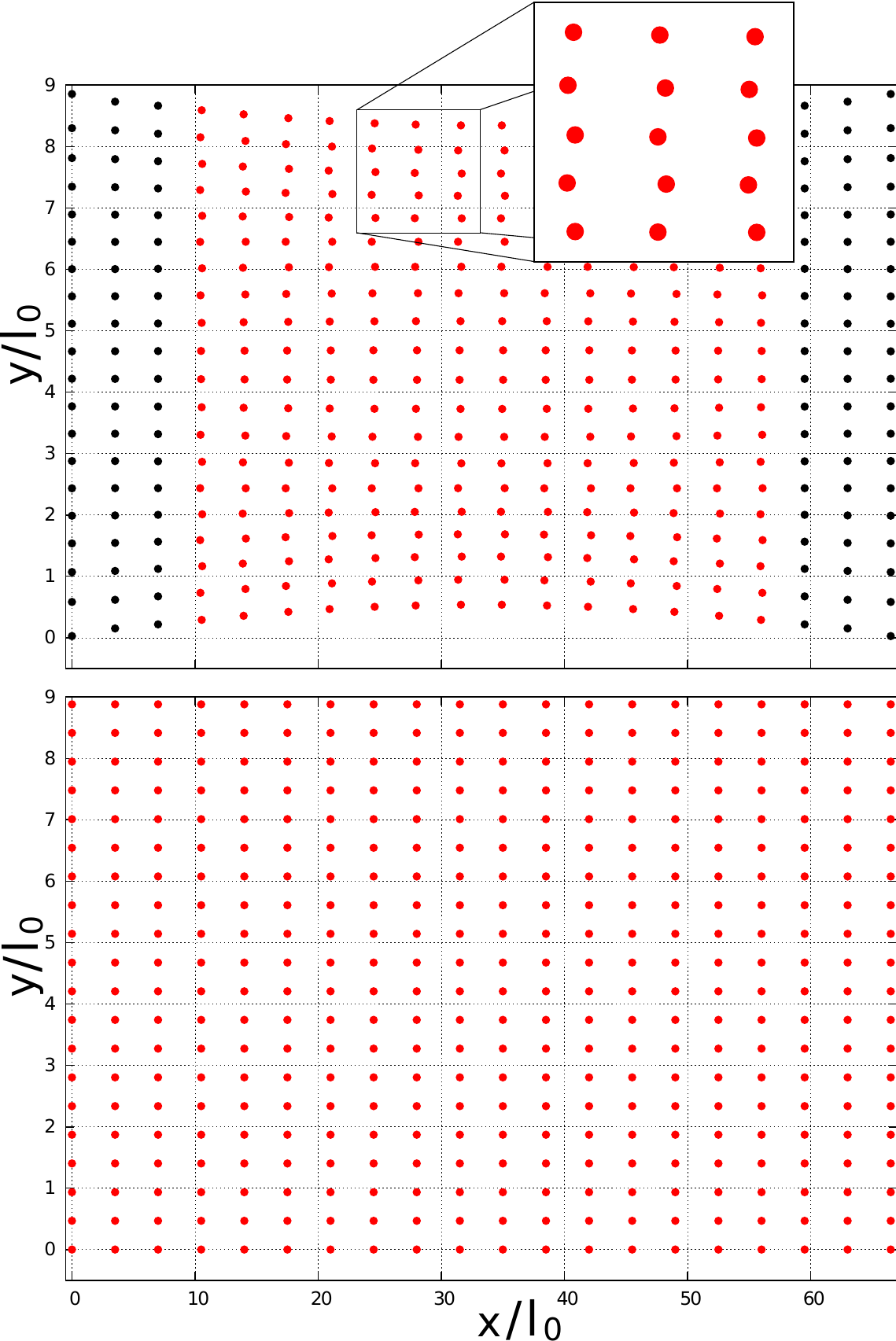}
    \caption{
     An initial square lattice undergoing the same total amount of horizontal strain
 \AM{at vanishing magnetic moment} and relaxed through NAT (top) and AT (bottom).
Clamped particles are colored in black in the NAT case.
The depicted deformations are much larger than the ones used in the following
to determine the elastic moduli (here the sample was expanded in $x$-direction by a factor of $2.5$).
    } 
  \label{at_nat_deform}
  \end{center}
\end{figure}

In a non-affine transformation (NAT), instead, most of the particles are free to adjust their positions
 independently of each other in 2D.
Only the particles on the two opposing edges of the sample are ``clamped'' and forced to move
 in a prescribed way along $x$-direction, but they are free to adjust in $y$-direction.
All clamped particles in the NAT are forced to be expanded in the $x$-direction in the same way as in the
 corresponding AT to allow better comparison. See Fig.~\ref{at_nat_deform} for an illustration of the two kinds
 of deformation.
To perform NAT minimization we have implemented the conjugated gradients algorithm
 \cite{hestenses1952methods,shewchuk1994introduction} using analytical expressions of the gradient
 and Hessian of the total energy.
\AM{Numerical thresholds were set such that the resulting error bars in the figures below are
 significantly smaller than the symbol size.}

As a consequence, NAT minimizes energy over $\simeq 2N$ degrees of freedom.
Since the NAT has many more degrees of freedom for the minimization than AT, we expect the former
 to always find a lower energetic minimum compared to the latter.
Thus, for the elastic modulus, we obtain $E^\AT\geq E^\NAT$.
Fig.~\ref{at_nat_deform} shows how NAT and AT minimizations yield different ground states for the same
 total amount of strain along $\xv$.

To compute the elastic modulus, we first find the equilibrium state through NAT for prescribed $\vm$.
Next, using AT, we impose a small shrinking/expansion and after the described AT minimization
 obtain $E^\AT$ via Eq.~(\ref{elmod}).
Then, starting from the NAT ground state again, we perform the same procedure using the NAT minimization
and thus determine $E^\NAT$.


\section{Results}

In the following we will briefly discuss the behavior of the elastic modulus in the limit of
 large systems.
Then, on the one hand, we will demonstrate that introducing a randomization in the lattices dramatically affects
 the performance of affine calculations.
On the other hand, \AM{we will investigate how} in each case structure and relative orientation of the
 nearest neighbors determine the trend of $E(m)$.

\subsection{Elastic Modulus for Large Systems}\label{largeN}

We run our simulations for lattices of $N_x=N_y$.
It is known that the total elastic modulus of two identical springs in series halves, whereas,
 if they are in parallel doubles, compared to the elastic modulus of a single spring.
In our case of determining the elastic modulus in $x$-direction, the total elastic modulus
 $E$ will be proportional to $N_y/N_x$. Thus, with our choice of $N_x=N_y$, it should not depend on $N$.
We will investigate the exemplary case of a rectangular or square lattice for $m=0$ to estimate the impact
 of finite size effects on our results, since a simple analytical model can be used to predict
 the value of $E$.

\begin{figure}
  \begin{center}
    \includegraphics[width=8.6cm]{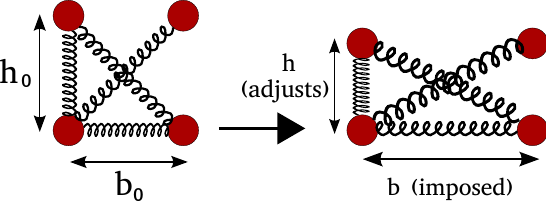}
    \caption{
Minimal rectangular model consisting of one $x$-oriented, one $y$-oriented, and two diagonal springs.
$b_0$ and $h_0$ are the base and height of the rectangular cell in the unstrained state.
Under strain, $b_0\rightarrow b$ and the height is free to adjust in order to
minimize the elastic energy, $h_0\rightarrow h$.
    } 
  \label{minimal_model}
  \end{center}
\end{figure}

Our rectangular lattice is made of vertical and horizontal springs coupling nearest neighbors
 and diagonal springs connecting next-nearest neighbors. 
The diagonal springs 
are necessary to avoid an unphysical soft-mode shear instability of the bulk rectangular crystal. 
In the large-$N$ limit there are on average one horizontal, one vertical, and two diagonal springs per particle.
The deformation of a corresponding ``unit spring cell'' is depicted in Fig.~\ref{minimal_model}.
$b_0$ and $h_0$ are, respectively, the length of the horizontal and vertical spring of the unit cell
 in the undeformed state, whereas $b$ and $h$ are the respective quantities in the deformed state.
$b$ is fixed by the imposed strain, whereas $h$ adjusts to minimize the energy, $\partial U/\partial h =0$.
This model describes, basically, the deformation of a cell in the bulk within an AT framework.

If magnetic effects are neglected, we find that the linear elastic modulus of such a system is
\begin{equation}\label{G0anal}
 E(m=0) \simeq \frac{d^2U_{el}}{{db}^2} \Biggr|_{b=b_0} = k\left( 1+ \frac{2 r_0^2}{3+r_0^2} \right).
\end{equation}
Here $r_0=b_0/h_0$ is the base-height ratio of the unstrained lattice. Furthermore, we have linearized the
$h(b)$ deformation around $b=b_0$.

\begin{figure}
  \begin{center}
    \includegraphics[width=8.6cm]{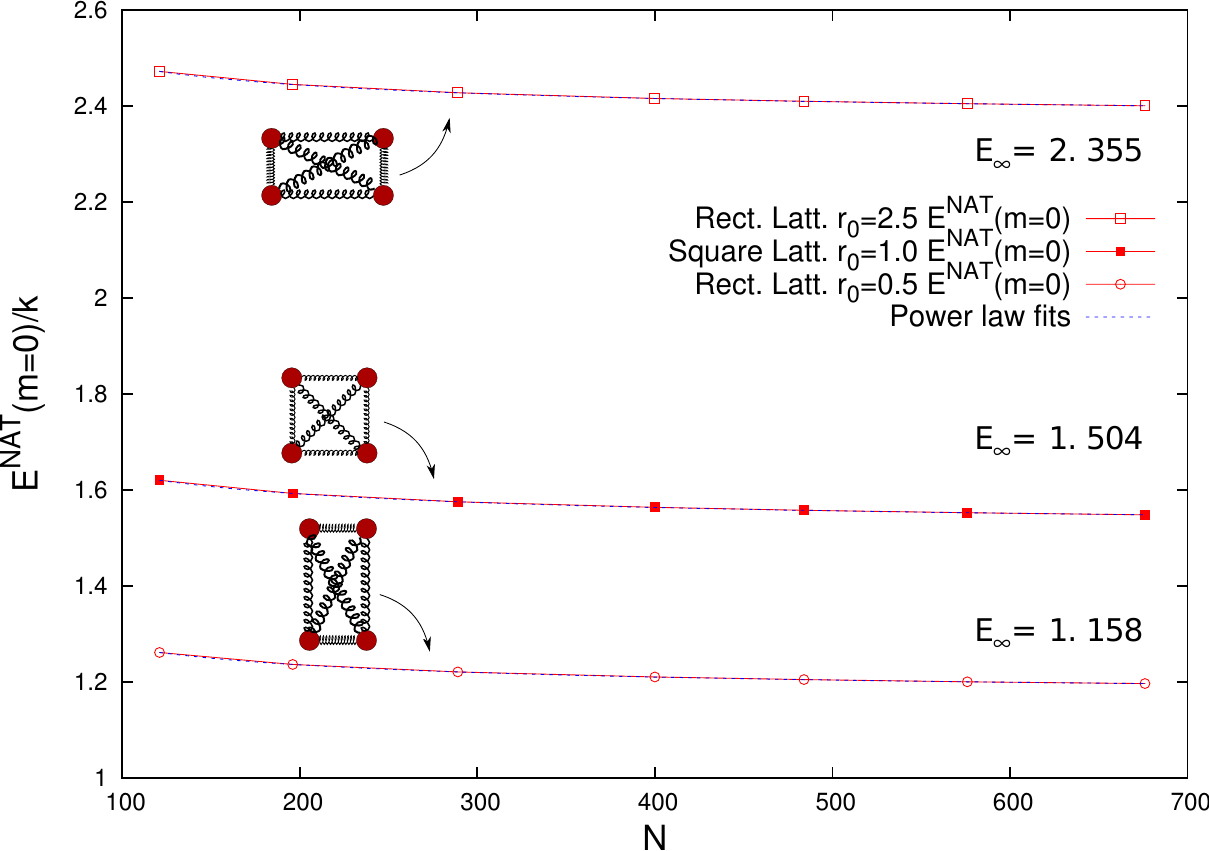}
    \caption{
   $E^\NAT(m=0)/k$ for different rectangular lattices increasing the number of particles $N$.
Fits with a power law of the form $E_N/k = E_\infty +\alpha N^\beta$ show a convergence towards
 the finite values indicated in the figure, while the values predicted by Eq.~(\ref{G0anal})
 are, (bottom to top curve) $1.154$, $1.500$, and $2.351$.
\AM{The values of $\beta$ resulting from the fit are (bottom to top curve) $-0.56$, $-0.55$, and $-0.54$.}
$E^\NAT(m\neq0)/k$ show the same convergence behavior for any $m$.
    } 
  \label{conv_np}
  \end{center}
\end{figure}

In the limit of large $N$, the elastic modulus determined by NAT should be dominated by bulk behavior.
\AM{For regular rectangular lattices stretched along the outer edges of the lattice cell,
 the deformation in the bulk becomes indistinguishable from an affine deformation.
We therefore can use our analytical calculation to test whether our systems are large enough
 to correctly reproduce the elastic modulus of the bulk.}
For this regular lattice structure it should correspond to the modulus following from Eq.~(\ref{G0anal}).
We calculated \AM{numerically} $E^\NAT(m=0)$ for different rectangular lattices
 as a function of $N$ and plot the results in Fig.~\ref{conv_np}.
Indeed, for large $N$, we find the convergence as expected.

From Fig.~\ref{conv_np} we observe that the modulus has mostly converged to its large-$N$ limit at $N=400$,
 therefore most of our calculations are performed for $N=400$ particles.
We have checked numerically that a similar convergence holds for any investigated choice of $\vm$ and
 lattice structure. 
For any $\vm\neq \mathbf{0}$ that we checked, we found a similar convergence behavior
 as the one depicted for $m=0$ in Fig.~\ref{conv_np}.

\subsection{Impact of Lattice Randomization on AT Calculations}

We have seen how, in the large-$N$ limit, AT analytical models and NAT numerical calculations converge
 to the same result in the case of regular rectangular lattices.
In fact, we expect AT to be a reasonable approximation in this regular lattice case, since it conserves
 the initial shape of the lattice.
For symmetry reasons, this behavior may be expected also for NAT at small degrees of deformation.
But how does AT perform in more realistic and disordered cases where the initial
 particle distribution can be irregular?
To answer this question we will consider the difference $E^\AT-E^\NAT$, the elastic modulus
 numerically calculated with AT and NAT, at $m=0$, for different and increasingly randomized lattices.

\begin{figure}
  \begin{center}
    \includegraphics[width=8.6cm]{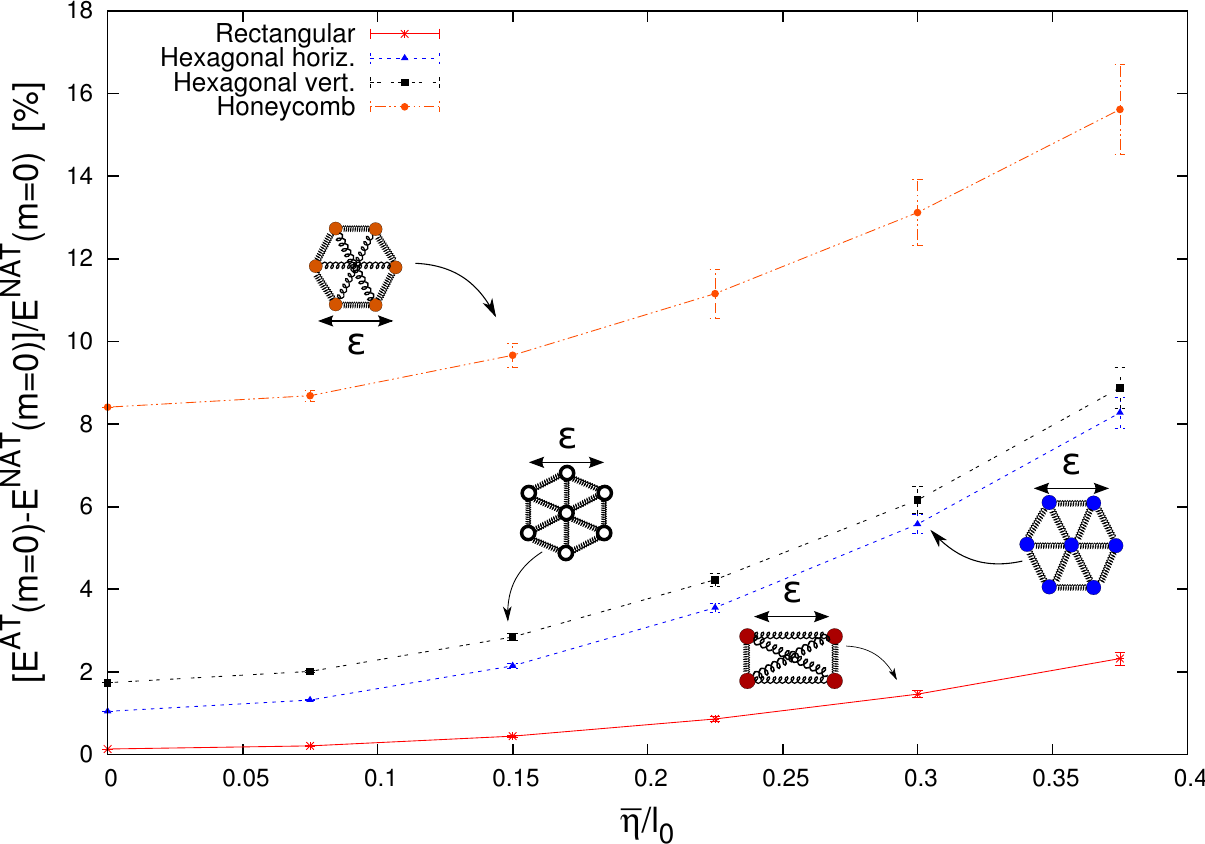}
    \caption{
 Different lattices for $m=0$ with the initial unstrained state randomized by displacing
each particle by $(\eta_1,\eta_2 )$, where $\eta_1,\eta_2$ are stochastic variables uniformly distributed in
$[-\overline{\eta}/2,\overline{\eta}/2]$.
\AM{Each point in this figure is obtained averaging over $100$ different realizations of random distributions
 generated from the same starting regular lattice.
Error bars were obtained from the resulting standard deviation.}
We indicate the direction of the applied strain by the double arrow marked by $\varepsilon$.
The difference in $E(m=0)$ calculated on the one hand by AT and on the other hand by NAT is plotted
 in $\%$ of $E^\NAT(m=0)$ as a function of an increasing randomization parameter $\overline{\eta}$.
    } 
  \label{G_x_rand}
  \end{center}
\end{figure}

We have considered a rectangular lattice with diagonal springs, a hexagonal lattice with horizontal rows
 of nearest neighbor springs, one with vertical rows, and a honeycomb lattice with springs beyond
 nearest neighbors (as depicted in Fig.~\ref{G_x_rand}).

To obtain the randomized lattices, we start from their regular counterparts and randomly move
 each particle within a square box of edge length $\overline{\eta}$ and centered in the regular lattice site.
We call $\overline{\eta}$ the randomization parameter used to quantify the degree of randomization.
\AM{In our numerical calculations we increased $\overline{\eta}$ up to $\overline{\eta}=0.375 l_0$.
This is an appreciable degree of randomization considering that at $\overline{\eta}= l_0$
 two nearest neighbors in a square lattice may end up at the same location.
To average over different realizations of the randomized lattices, we have performed $100$ numerical runs
 for every initial regular lattice and every chosen value of $\overline{\eta}$.}
 In Fig.~\ref{G_x_rand}, we plot the relative difference between $E^\AT$ and $E^\NAT$. 

Already for the regular lattices of vanishing randomization $\overline{\eta}=0$, we find a relative deviation
 of $E^\AT$ from $E^\NAT$ in the one-digit per-cent regime.
This deviation is smallest for the regular rectangular lattice, where the principal stretching directions
 are parallel to the nearest-neighbor bond vectors.
The deviation for $\overline{\eta}=0$ increases when we consider instead the hexagonal and honeycomb lattices. 
Obviously, and this is our main point here, the relative difference between $E^\AT$ and $E^\NAT$ increases
 for each lattice that we investigated  with the degree of randomization $\overline{\eta}$.
Therefore NAT find much lower equilibrium states with increasing randomization, and AT lead to erroneous results.
\AM{So far, however, we could not yet establish a simple rigid criterion that would quantitatively predict
 the observed differences between AT and NAT.}

\subsection{The case $\vm \parallelsum \xv$}

We will now consider a non-vanishing magnetic moment $\vm \parallelsum \xv$. 
This is parallel to the direction in which we apply the strain in order to measure the elastic modulus. 
As we will see below, the behavior of the elastic modulus as a function of the magnetic moment $E(m)$
 strongly depends on the orientation of $\vm$ and on the lattice structure.
The kind of magnetic interaction between nearest neighbors is fundamental for its impact on the elastic modulus.
On the one hand, when the magnetic coupling between two particles in $U_m$ [see Eq.~(\ref{emagn})] is solely
 repulsive, i.e.\ $\vm \perp \bm{r}_{ij}$, its second derivative is positive and therefore gives a positive
 contribution to the elastic modulus.
On the other hand, when  $\vm \parallelsum \bm{r}_{ij}$ the interaction is attractive and the second derivative
 of $U_m$ gives a negative contribution to the elastic modulus.

\begin{figure}
  \begin{center}
    \includegraphics[width=8.6cm]{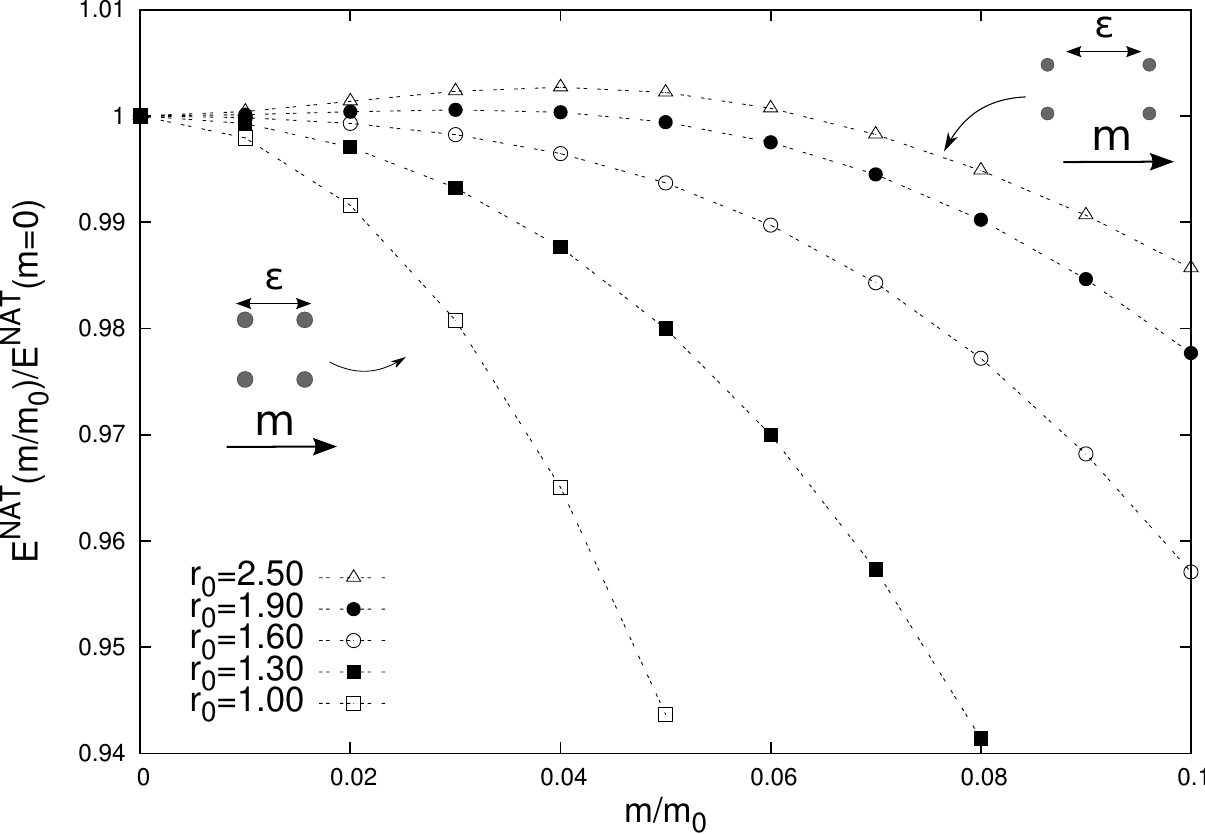}
    \caption{
      Rectangular lattice with $\vm \parallelsum \xv$. Different trends of $E^\NAT(m/m_0)$ are shown for
different unstrained lattice shapes using the undeformed base-height ratio $r_0$ as shape parameter.
We indicate the direction of the applied strain by $\varepsilon$.
To compare and enhance the different trends, $E^\NAT(m/m_0)$ is rescaled by $E^\NAT(m=0)$.
    } 
  \label{G_x_rect}
  \end{center}
\end{figure}

When $\vm$ is parallel to the strain direction $\xv$, the magnetic interaction along $\xv$ is attractive and,
 for $m$ large enough, will cause the lattice to shrink and the elastic modulus to decrease.
For some cases, though, $E(m)$ shows an initial increasing trend.
This happens when in the unstrained lattice the particles are much closer in $\yv$ than in $\xv$.
Then, for small deformations, magnetic repulsion is prevalent and the magnetic contribution to $E$
 is positive, as can be seen \AM{for the rectangular case} from Fig.~\ref{G_x_rect}.

The total energy of the system is the sum of elastic and magnetic energies. Since the derivative is a linear
 operator, the elastic modulus can be decomposed in elastic and magnetic components: $E=E_{el}+E_m$.
The analytical calculation for the minimal rectangular system described in subsection \ref{largeN}
 applied to this configuration and considering magnetic interaction up to nearest neighbors only,
 predicts that
\begin{equation}\label{Gmxanal}
 E_m \simeq \frac{d^2U_{m}}{{db}^2}\Biggr|_{b=b_0} = \frac{12 m^2}{b_0^5}\left( -2+ \frac{4 r_0^7}{{(3+r_0^2)}^2} \right)
\end{equation}
in the rectangular case.

From Eq.~(\ref{Gmxanal}) we expect a magnetic contribution to the total elastic modulus
 increasing  with $m$ for $r_0\geq 1.175$ and decreasing with $m$ for $r_0\leq 1.175$.
Qualitatively we observe this trend for $m/m_0\ll 1$ in Fig.~\ref{G_x_rect}.
However, it seems that the the initial
 trend for $E(m)$, i.e.\ close to the unstrained state, switches from increasing to decreasing
 around $r_0\simeq 1.60$, higher than we expected.
Although the minimal analytical model can predict the existence of a threshold value for $r_0$ it would
 need the magnetic contribution of more than only nearest neighbor particles to be more accurate,
 since the magnetic interaction is long ranged (whereas the elastic interaction acts only on nearest neighbors).

\subsection{The case $\vm \parallelsum \yv$}

In this orientation of the magnetic moment, the hexagonal lattice case is exemplary, because it shows very well
 the orientational structural dependence of $E(m)$.

\begin{figure}
  \begin{center}
    \includegraphics[width=8.6cm]{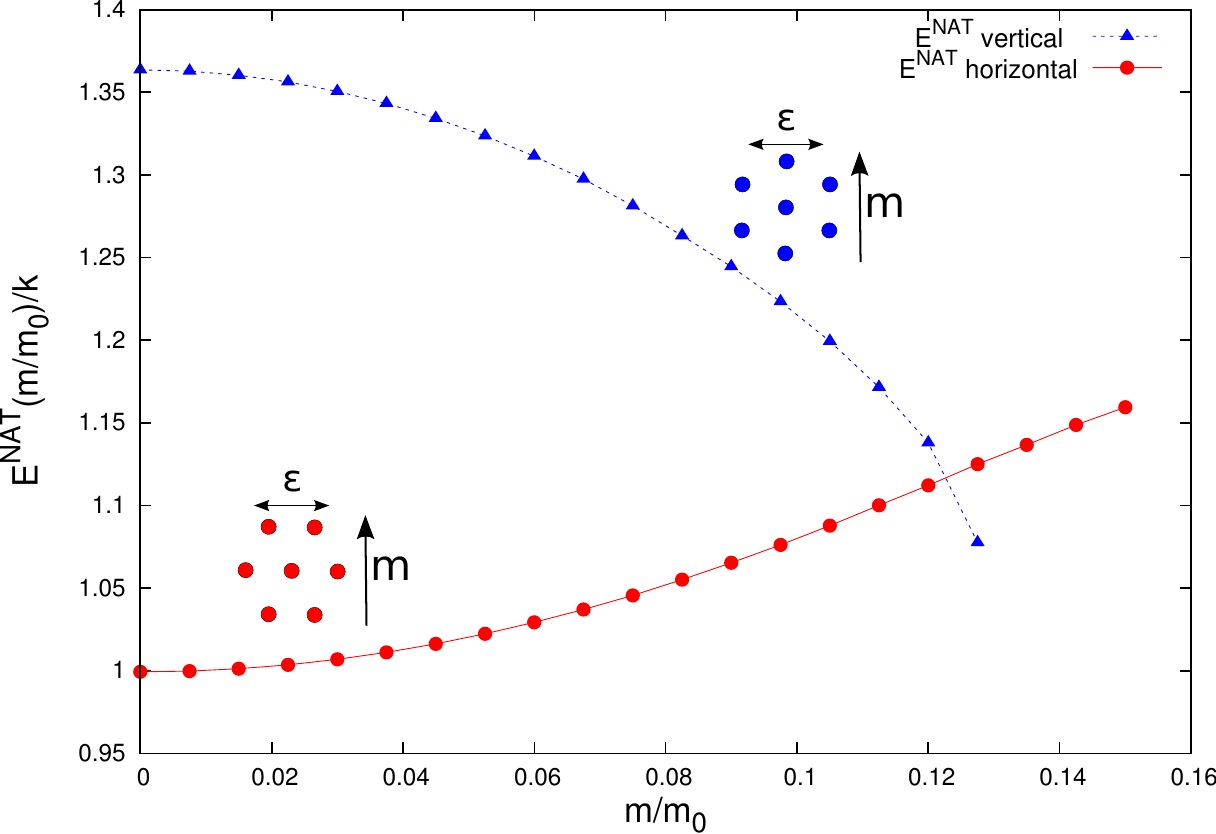}
    \caption{
      Hexagonal lattice with $\vm \parallelsum \yv$ for a hexagonal lattice with horizontal rows
 (bottom inset, where two nearest neighbors are oriented along $\xv$) and for one with vertical rows
 (top inset, where two nearest neighbors are oriented along $\yv$).
We indicate the direction of the applied strain by $\varepsilon$.
It is remarkable how the magnetic interaction between nearest neighbors and the $E^\NAT(m)$ behavior change
  when the lattice is rotated by $\pi/2$.
    } 
  \label{G_y_hex}
  \end{center}
\end{figure}

On the one hand, for the hexagonal lattice ``horizontally'' oriented (see the bottom inset in Fig.~\ref{G_y_hex})
 there are no nearest neighbors in the attractive direction $\yv$;
 there are instead two along $\xv$ whose interaction is purely repulsive, therefore the second derivative
 of their interaction $U_m$ is positive.
On the other hand, for the same lattice rotated by $\pi/2$ (see the top inset in Fig.~\ref{G_y_hex})
 there are two nearest neighbors in the direction of $\vm$ and their interaction is strongly attractive;
 therefore, the second derivative of their interaction $U_m$ is negative.

The result, as can be seen in Fig.~\ref{G_y_hex}, is that in the former case the elastic modulus
is increasing and in the latter is decreasing.

\subsection{The case $\vm \parallelsum \zv$}

\begin{figure}
  \begin{center}
    \includegraphics[width=8.6cm]{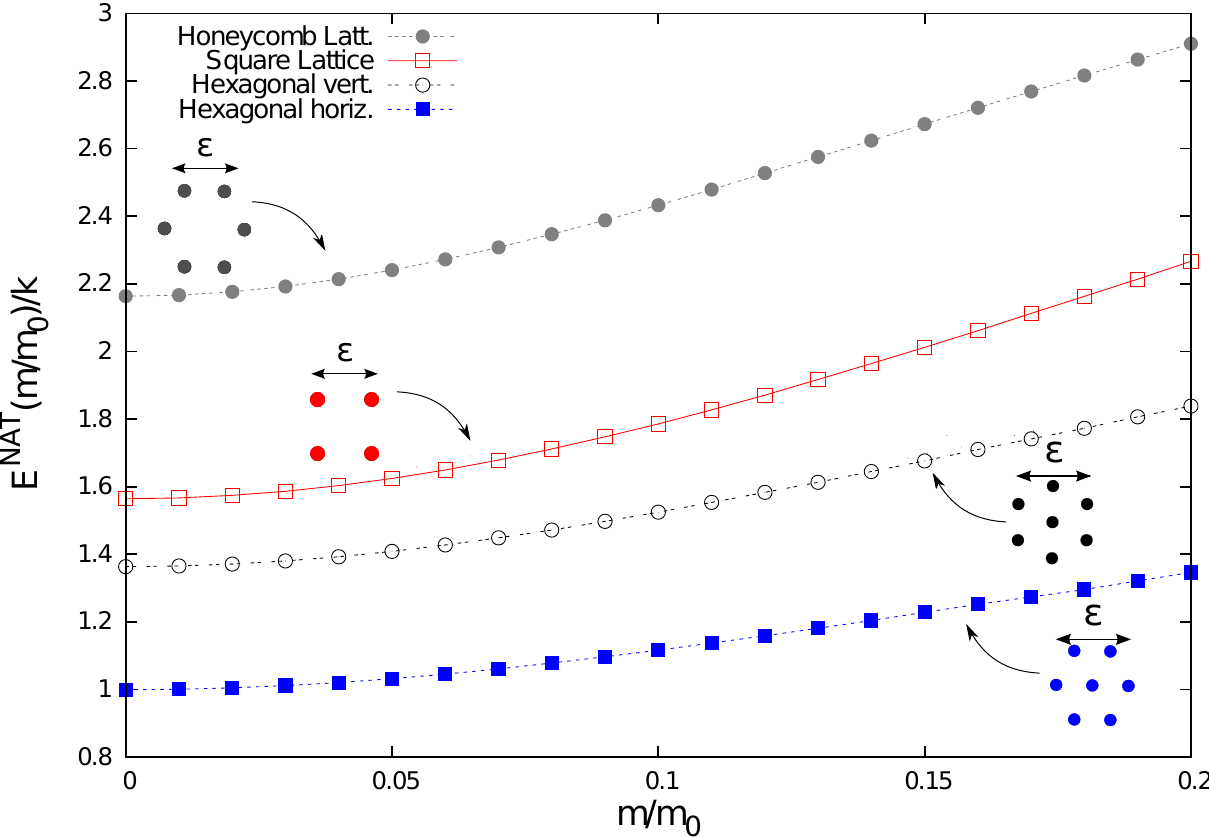}
    \caption{
Elastic modulus $E(m/m_0)/k$ calculated with NAT for $\vm \parallelsum \zv$ for the different
 lattices shown.
We indicate the direction of the applied strain by $\varepsilon$.
The magnetic interaction is purely repulsive and strengthens the elastic modulus in this configuration.
    } 
  \label{G_z_all}
  \end{center}
\end{figure}

In this configuration, the magnetic interactions between our particles are all repulsive
 and have the form $m^2/{r_{ij}}^3$.
The second derivative of the magnetic interparticle energy is always positive along the direction connecting
 the particles.
Therefore we expect the elastic modulus to be enhanced with increasing $m$, and $E(m)$ to be a monotonically
 increasing function.
As can be seen from Fig.~\ref{G_z_all}, this is true for all the different lattices we have considered.

\begin{figure} 
  \begin{center}
    \includegraphics[width=8.6cm]{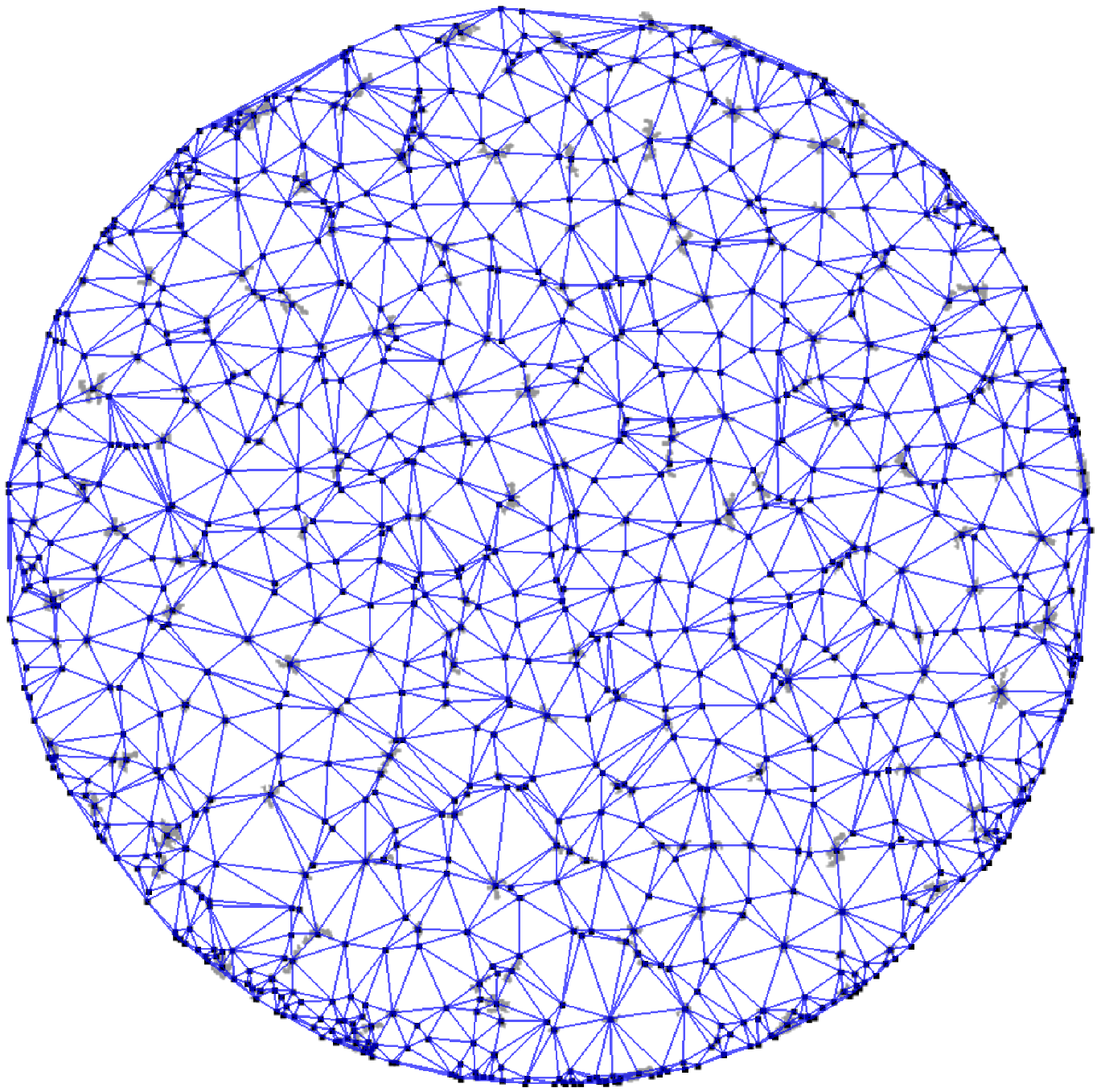}
    \caption{
Realistic lattice used to determine the elastic modulus as a function of the magnetic interactions
 in the case $\vm \parallelsum \zv$. 
The lattice was determined from an X-ray micro-tomographic image of a real experimental
 sample \cite{gunther2012xray} in the following way.
The sample was of cylindrical shape with a diameter of approximately $3$~cm. We show a cross-sectional
 cut through the sample at intermediate height. Inside the sample, the magnetic particles formed chains
 parallel to the cylinder axis, i.e.\ perpendicular to the depicted plane.
The average size of the particles was around $35$~$\mu$m.
Gray areas correspond to the tomographic spots
 generated by the magnetic particles in the sample and were identified by image analysis.
In our model, we then used the centers of these spots, marked by the black boxes,
 as lattice sites.
One magnetic particle was placed on each lattice site. Then the whole plane was tessellated
 by Delaunay triangulation with the particle positions at the vertices of the resulting triangles.
Elastic springs were set along the edges of the triangles. 
  [The micro-tomography data are taken from Ref.~\cite{gunther2012xray}, Fig.~5 (H=3~mm),
 \copyright~IOP Publishing. Reproduced by permission of IOP Publishing. All rights reserved.]
    } 
  \label{fig_sample}
  \end{center}
\end{figure}

We have already seen in Fig.~\ref{G_x_rand} how the randomization of the lattice seriously affects
the difference between AT and NAT. For the $\vm \parallelsum \zv$ case we have also considered
a real particle distribution taken from an experimental sample \cite{gunther2012xray}.
The real sample was of cylindrical shape with a diameter of about $3$~cm. It had the magnetic particles
 arranged in chain-like aggregates parallel to the cylinder axis and spanning the whole sample.
The positions of the particles were obtained through X-ray micro-tomography and subsequent image analysis.
We extracted the data from a circular cross-section taken approximately at half height of the cylinder
 and shown in Fig~\ref{fig_sample}.
In this way we consider by our model the physics of one cross-sectional plane of the cylindrical sample. 

\begin{figure} 
  \begin{center}
    \includegraphics[width=8.6cm]{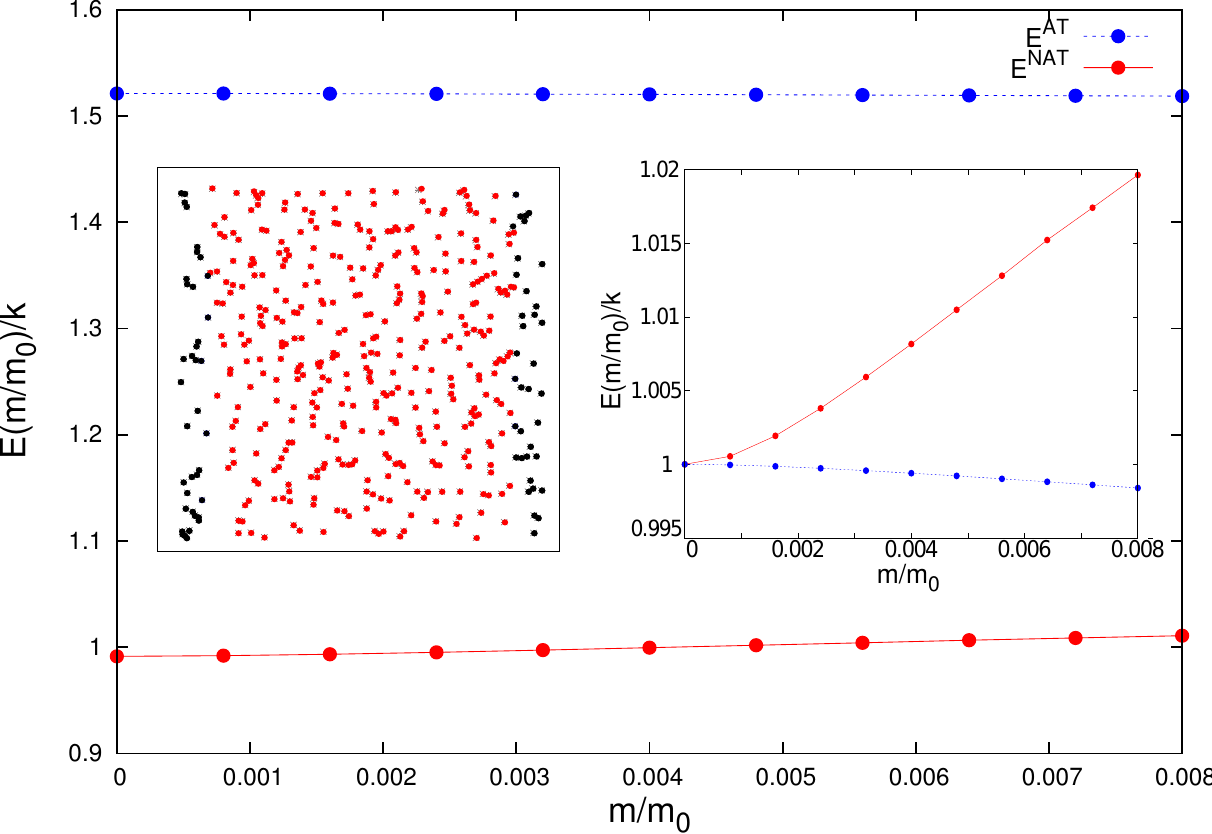}
    \caption{
Elastic modulus $E(m/m_0)/k$ calculated for $\vm \parallelsum \zv$ with NAT and AT techniques
 for the experimental lattice drawn in the left inset.
Black dots represent clamped particles.
Besides considerably overestimating the elastic modulus,  $E^\AT(m/m_0)/k$ shows a flat/decreasing
 behavior, whereas $E^\NAT(m/m_0)/k$ is correctly increasing.
In the right inset, we rescaled $E(m/m_0)$ by $E(m=0)$ to better show the two different trends.
\AM{The numerical error bars are much smaller than the symbol size.}
    } 
  \label{G_z_exp}
  \end{center}
\end{figure}

The extracted lattice was used as an input for our dipole-spring model. We placed a magnetic
 particle at the center of each identified spot in the tomographic image, see Fig~\ref{fig_sample}.
Guided by the situation in the real sample, the magnetic moments of the particles are chosen perpendicular
 to the plane (i.e.\ ``along the cylinder axis'').
The springs in the resulting lattice are set using Delaunay triangulation \cite{delaunay1934sur,
cgal:pt-t3-14a,borbath2012xmuct}
 with the particles at the vertices of the triangles and the springs placed at their edges.
Then, we cut a square block from the center of the sample containing the desired
 number of particles.
The clamped particles are chosen in such a way that they cover about $10\%$
 of the total area (see left inset in Fig.~\ref{G_z_exp}).

Again we numerically investigate two-dimensional deformations within the resulting two-dimensional layer.
If, in the future, this is to be compared to the case of a real sample, the deformations of this sample
 in the third direction, i.e.\ the anisotropy direction, have to be suppressed. For instance, the sample
 could be confined at the base and cover surfaces and compressed along one of the sides.
Then it can only extend along the other side.
Thus, within each cross-sectional plane, an overall two-dimensional deformation occurs, with macroscopic
 deformations suppressed in the anisotropy direction. 

As we can see from Fig.~\ref{G_z_exp}, in our numerical calculations for this case, AT  leads to a serious
 overestimation of the elastic modulus compared to the one obtained for NAT.
Moreover, as can be seen in the right inset of Fig.~\ref{G_z_exp}, the former predicts an erroneous
 flat/decreasing trend for $E(m)$, whereas the latter shows instead a correct increasing behavior.
This result can be interpreted considering that in AT all the particles must move
 in a prescribed way along each direction.
When the particle arrangement is irregular, some couples are very close and some are very distant. 
The erroneous trend in AT is mainly attributed to the very close particle pairs.
AT can force them to still move closer together despite the magnetic repulsion, whereas NAT
 allows them to avoid such unphysical approaches. 
Therefore, in order to properly minimize the energy, each particle must be free to adjust position
 individually with respect to its local environment.
As a consequence, for such realistic lattices AT provide erroneous results both quantitatively and
 qualitatively, making NAT mandatory in most practical cases.

\begin{figure} 
  \begin{center}
    \includegraphics[width=8.6cm]{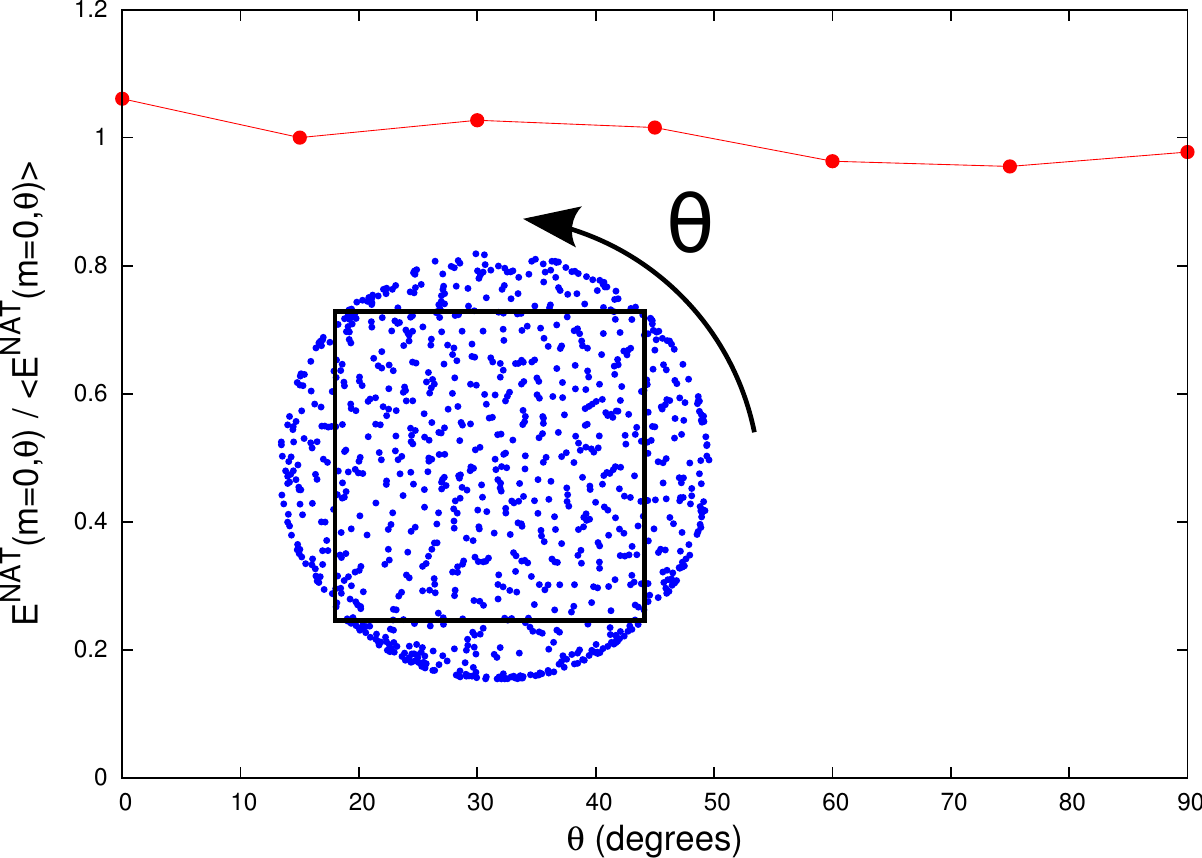}
    \caption{
Zero-field elastic modulus $E^\NAT(m=0)$ calculated with NAT for the experimental lattice
 drawn in the inset picture varying the rotation angle $\theta$.
To illustrate the effective isotropy we plot the elastic modulus rescaled by the average
 of $E^\NAT(m=0)$ over $\theta$.
The black square in the inset contains the block of particles extracted from the experimental data
 after the rotation and used in our calculation.
    } 
  \label{G_rot_exp}
  \end{center}
\end{figure}

Since within the analyzed two-dimensional cross-sectional layer the particle distribution appears to be rather
 isotropic, we expect the elastic modulus to be approximately the same in any direction in the plane.
To demonstrate this fact, we rotate the configuration in the plane with respect to the stretching direction
 by different angles $\theta$ between $0$ and $\pi/2$.
\AM{As we can see from Fig.~\ref{G_rot_exp}, the zero-field elastic modulus $E(m=0)$ shows only
 small deviations for the different orientations.
The origin of such deviations is ascribed to the square-cutting procedure which, after a rotation
 by an angle $\theta$, produces samples containing different sets of particles, each with different
 local inhomogeneities in particle distribution and spring orientation.}
\AM{For samples large enough to significantly average over all these different local inhomogeneities,
 the angular dependence of $E(m=0)$ should further decrease.
We found that for any rotation angle $\theta$, the behavior of $E(m)$ is similar to the one in
 Fig.~\ref{G_z_exp} corresponding to $\theta=0$, supporting our statement about the erroneous AT
 result.}

\subsection{Shear modulus}
\AM{For the set-up described in the previous subsection (see the left inset
 of Fig.~\ref{G_z_exp} with $\vm \parallelsum \zv$) we have also calculated the shear modulus $G(m)$
 as a function of the magnetic moment, for both AT and NAT.
The shear modulus is defined as the second derivative of the total energy $U$ with respect to a small
 displacement $\delta_y$ of the clamps in $y$-direction:
\begin{equation}\label{shmod}
 G = \frac{d^2 U}{{d\delta_y}^2} \simeq \frac{U(-\delta_y)+U(\delta_y)-2U(0)}{{\delta_y}^2}.
\end{equation}
In this calculation, to allow for the comparison between the results from AT and NAT, all particles
 within the clamped regions are forced to move in a prescribed (affine) way.}

\begin{figure} 
  \begin{center}
    \includegraphics[width=8.6cm]{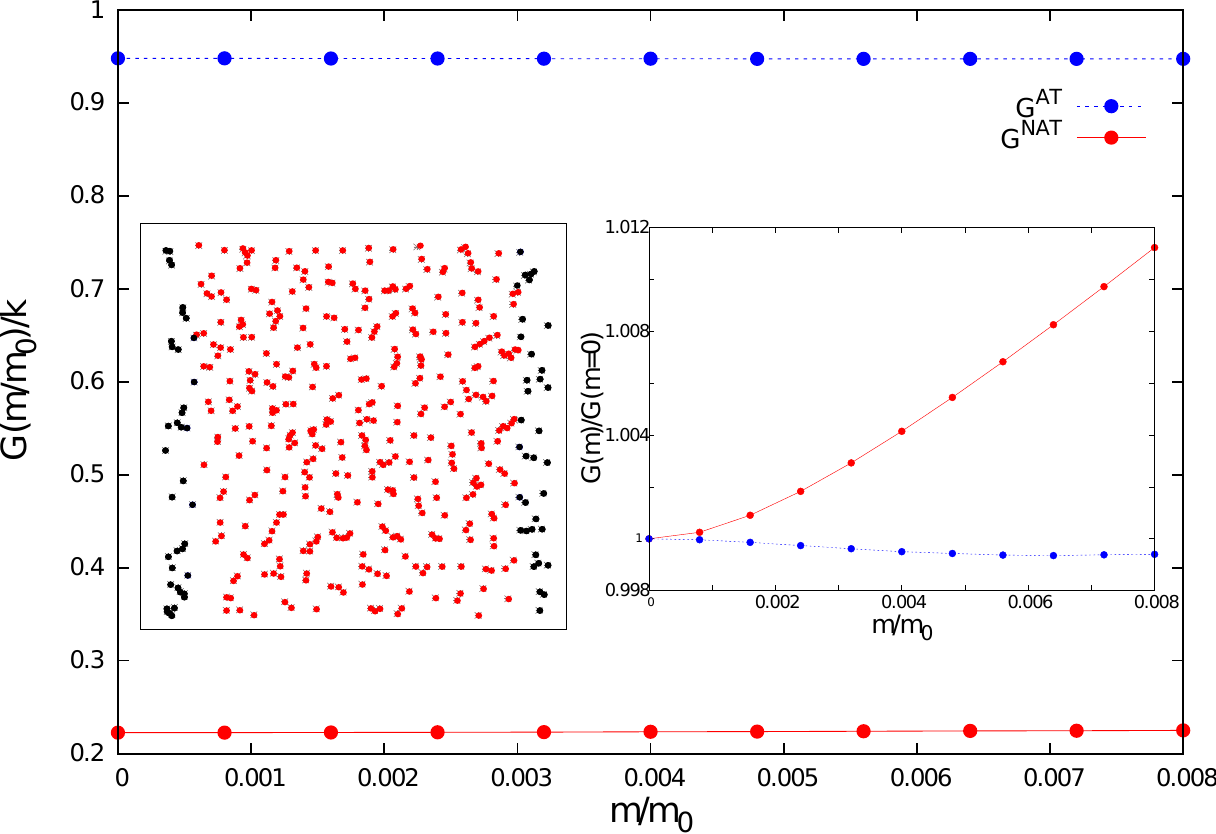}
    \caption{
\AM{Shear modulus $G(m/m_0)/k$ calculated for $\vm \parallelsum \zv$ with NAT and AT techniques
 for the experimental lattice drawn in the left inset.
Black dots represent clamped particles.
Here again, besides considerably overestimating the elastic modulus,  $G^\AT(m/m_0)/k$ shows a
 flat/decreasing behavior, whereas $G^\NAT(m/m_0)/k$ is correctly increasing.
In the right inset, we rescaled $G(m/m_0)$ by $G(m=0)$ to better show the two different trends. The numerical error bars are much smaller than the symbol size.}
    } 
  \label{Sh_z_exp}
  \end{center}
\end{figure}

\AM{It turns out that the behavior of the shear modulus is qualitatively the same as for the compressive
 and dilative elastic modulus (see Fig.~\ref{Sh_z_exp}).
Again, an incorrect decreasing behavior for the AT calculation is obtained.
In numbers, the relative difference between the AT and NAT results is larger than for the compressive
 and dilative elastic modulus.
Here we set $\delta_y$ as one percent of the dimension of the sample.
In Fig.~\ref{Sh_z_exp}, this choice produces numerical error bars much smaller than the symbol size.}

\section{Conclusions}

We have shown how the induction of aligned magnetic moments can weaken or strengthen the elastic modulus
 of a ferrogel or magnetic elastomer  according to lattice structure and nearest-neighbor orientations.
The orientation of nearest neighbors plays a central role.
If the vector connecting two nearest neighbors lies parallel to the magnetic moment,
 they attract each other, the second derivative of their magnetic interaction is negative, and the
 corresponding contribution to the total elastic modulus is negative, too.
If, instead, the nearest neighbors lie on a direction perpendicular to the magnetic moment, the
 second derivative of their magnetic interaction is positive and it tends to increase the total elastic modulus.
This effect can be seen modifying the nearest-neighbor structure, for instance tuning the shape of a rectangular
 lattice or rotating a hexagonal lattice.
We have also seen how the performance of affine transformations worsens for randomized
 and more realistic particle distributions, making non-affine transformation calculations mandatory when working
 with data extracted from experiments.

In the present case, we scaled out the typical particle separation and the elastic constant from the
 equations to keep the description general.
Both quantities are available when real samples are considered.
The mean particle distance follows from the average density, while the elastic constant could be
 connected to the elastic modulus of the polymer matrix.

The dipole-spring system we have considered is a minimal model. We look forward to improving it in
 different directions. First, we would like to go beyond linear elastic interactions using non-linear springs,
 perhaps deriving a realistic interaction potential from experiments or more microscopic simulations.
Second, the use of periodic boundary conditions may improve the efficiency of our calculations
 and give us new insight into the system behavior (although we demonstrated by our study of asymptotic
 behavior that border effects are negligible in the present set-up).
Furthermore, we may include a constant volume constraint, since volume conservation is not rigidly enforced
 in the present model. 
To isolate the effects of different lattice structures and the assumption of affine deformations,
 we here assumed that all magnetic moments are rigidly anchored along one given direction.
 In a subsequent step, this constraint could be weakened by explicitly implementing the interaction
 with an external magnetic field or an orientational memory. 
Finally, to build the bridge to real system modeling, an extension of our calculations to three dimensions
 is mandatory in most practical cases.

\begin{acknowledgments}
The authors thank the Deutsche Forschungsgemeinschaft for support of this work through the priority
 program SPP 1681.
\end{acknowledgments}


\end{document}